\documentclass[aps,prl,reprint,amsmath,amssymb]{revtex4-2}
\pdfoutput=1
\usepackage{xcolor, soul}
\sethlcolor{yellow}
\usepackage{xcolor-patch}
\usepackage{graphicx}
\usepackage{dcolumn}
\usepackage{bm}
\usepackage{bbold}
\usepackage{natbib}
\usepackage{braket}
\usepackage{mathtools}
\usepackage[breaklinks=true,colorlinks,citecolor=blue,linkcolor=blue,urlcolor=blue]{hyperref}
\begin{document}
\newcommand{\defeq}{\vcentcolon=}
\newcommand{\around}{{\raise.17ex\hbox{$\scriptstyle\sim$}}}

\title{First-principles phase diagram of an interacting ionic chain}
\author{Jamin Kidd}
\author{Ruiqi Zhang}
\author{Shao-Kai Jian}
\author{Jianwei Sun}
    \email{jsun@tulane.edu}
\affiliation{\mbox{Department of Physics and Engineering Physics, Tulane University, New Orleans, Louisiana 70118, USA}}

\date{\today}	

	\begin{abstract}
		The widely-used Kohn-Sham implementation of density functional theory (DFT) maps a system of interacting electrons onto an auxiliary non-interacting one and is presumably inaccurate for strongly correlated materials. We present a concrete benchmark for DFT by examining the electronic ground state phase diagram of a strongly interacting chain with uneven, non-integer nuclear charges. The interplay between charge imbalance and Coulomb repulsion yields two competing phases, a band insulator and a Mott insulator. By including infinitesimal lattice distortions, DFT stabilizes the intermediate spontaneously dimerized insulator phase that results from this competition. We assess the phase diagram by mapping the bond length and nuclear charge ratio of the chain to the ionic Hubbard model and performing highly accurate density matrix renormalization group calculations. Our comparative study provides foundational insight into the utility of symmetry-broken DFT as a predictive tool for elucidating phase diagrams with competing orders.
	\end{abstract}

	\maketitle


    \textit{Introduction.}---Understanding the exotic phases of strongly correlated materials (like high-$T_c$ cuprate superconductors \cite{high-tc}) remains a central challenge in condensed matter physics, primarily owing to the lack of an exact solution for the requisite many-body electronic Hamiltonian. Among the various approximate methods used to investigate material candidates, Kohn-Sham density functional theory (DFT) \cite{kohn-sham,yang-dft} is often chosen for its balance between accuracy and computational efficiency. However, conventional DFT (which includes both the local-spin-density approximation \cite{lsda1,lsda2} and the generalized gradient approximation \cite{gga}) typically yields poor results in the strong interaction regime, wherein self-energy plays an important role \cite{pickett}. Interestingly, our recent work has shown that the combination of improved functionals \cite{scan1,scan2} and explicit symmetry breaking \cite{polymorphous,dft-sym-break,ZungerFeSe} significantly enhances the DFT description of several well-known correlated systems, including cuprates \cite{chrisLCO, scan-cuprate, ybco}, nickelates \cite{ruiqi_nickelate}, and rare-earth compounds \cite{smb6}. A baseline comparison of this symmetry-broken methodology to exact theory could illuminate the fundamental limitations of DFT but has been explored very little to date. 

\begin{figure}
        \centering
        \includegraphics[width=\linewidth]{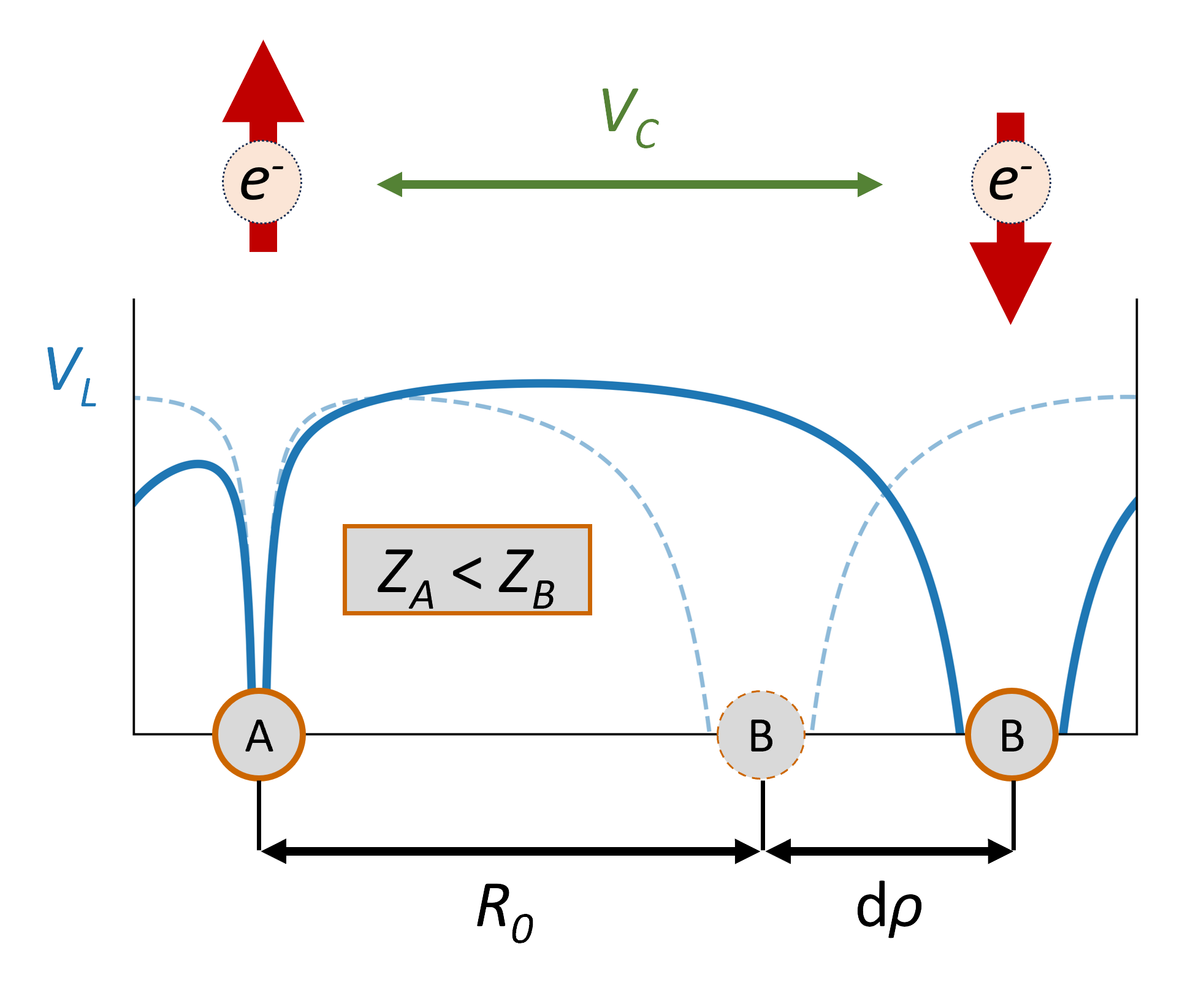}
        \caption{Schematic depicting the degrees of freedom considered in the DFT simulation of the NNC chain (only one unit cell is shown). $V_L$ is the lattice potential (Eqn. \ref{eqn:lattice-potential}), whose shape is determined by $R_0$, $\mu = Z_A/Z_B$, and $\text{d}\rho$ (the dashed line is the case $\text{d}\rho=0$). $V_C$ is the Coulomb repulsion (Eqn. \ref{eqn:coulomb}), which is long-range in DFT.}
        \label{fig:schematic}
    \end{figure}

    In this Letter, we use DFT to study 
    the rich phase diagram of a simple one-dimensional (1D) system and validate it using exact many-body techniques. In particular, we consider a chain whose unit cell hosts two evenly spaced nuclei, A and B, and two electrons. The nuclei are allowed to have uneven, non-integer charge numbers $Z_A$ and $Z_B$, respectively, and we restrict their sum to $+2$ so that the cell is neutral. Subsequently, we refer to this chain as the non-integer nuclear charge (NNC) chain. Our DFT simulations reveal band insulator (BI) and Mott insulator (MI) phases in the regimes dominated by charge imbalance and Coulomb interaction, respectively. More strikingly, by including an infinitesimal lattice distortion $\text{d}\rho$, we find that an intermediate spontaneously dimerized insulator (SDI) appears as a result of the strong competition. While DFT has previously been used to investigate Mott physics in the 1D hydrogen chain \cite{h-chain1,h-chain2}, this newly discovered SDI phase emerges from the additional degrees of freedom considered in our calculations, which are illustrated in Fig. \ref{fig:schematic}. We benchmark the phase diagram against a highly reliable density matrix renormalization group (DMRG) simulation \cite{dmrg} by mapping the NNC chain to the ionic Hubbard model (IHM) \cite{ihm-organic,ihm-cuprate}. The mapping, detailed below, captures the essential physics of the NNC chain, and the independent DMRG calculations confirm the SDI phase. It should be noted that the phase diagram of the IHM is well understood \cite{ihm-bosonization,ihm-mtt,ihm-dmrg} and hence provides an accurate benchmark. The consistency between DFT and DMRG is evidence that explicit symmetry breaking complements DFT as a method for exploring the phase diagram of strongly correlated materials.

\begin{figure}[b]
    \centering
    \includegraphics[width=\linewidth]{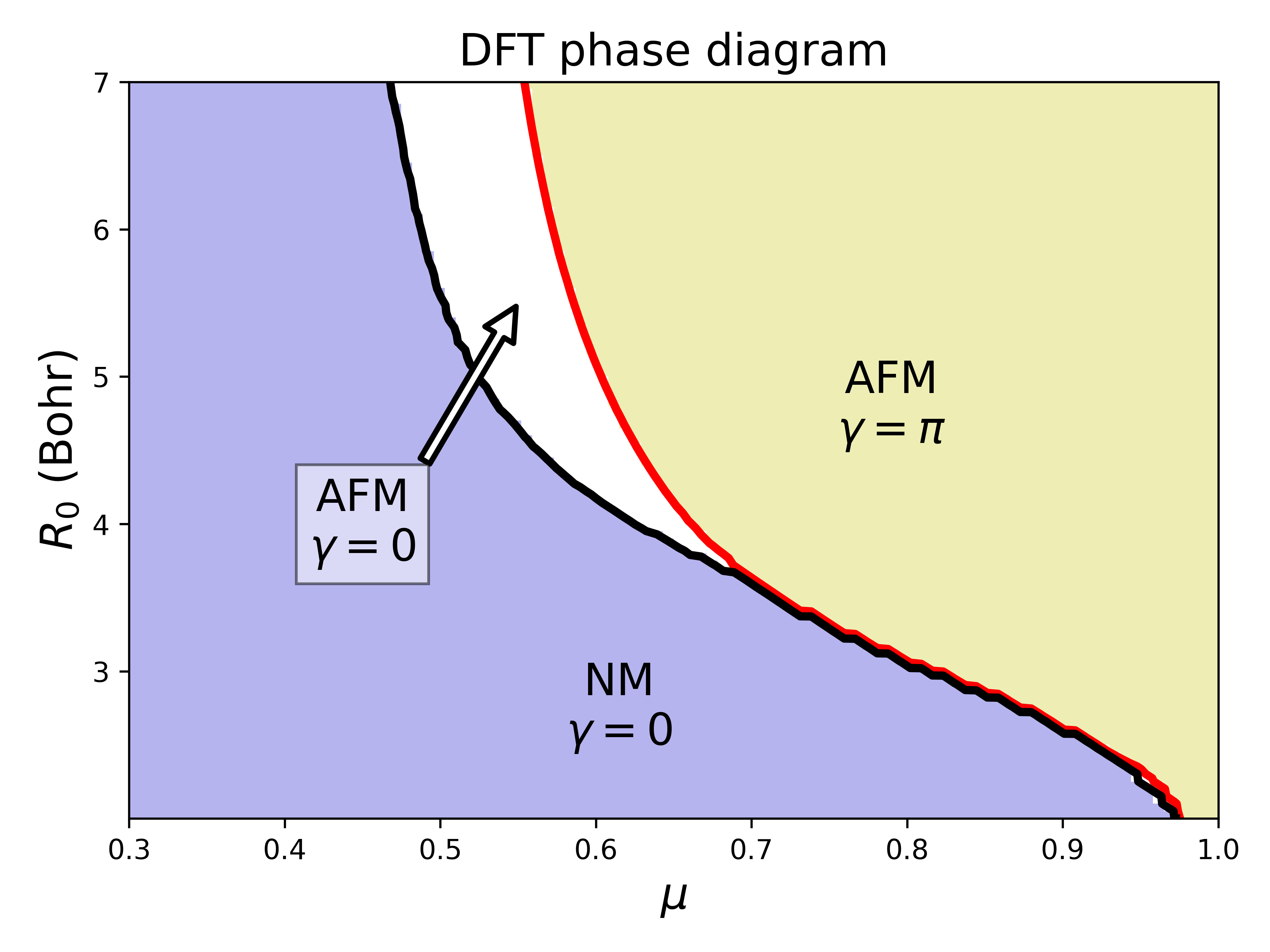}
    \caption{Ground state magnetic ordering and Berry phase $\gamma$ of the NNC chain as functions of $R_0$ and $\mu$ for $\text{d}\rho = 0$. The chain is antiferromagnetic (AFM) to the right of the black curve and nonmagnetic (NM) to the left. To the right (left) of the red curve, the Berry phase is $\pi$ (zero), and the band gap closes along this boundary.}
    \label{fig:dft-phase-diagram}
\end{figure}

\begin{figure}
        \centering
        \includegraphics[width=\linewidth]{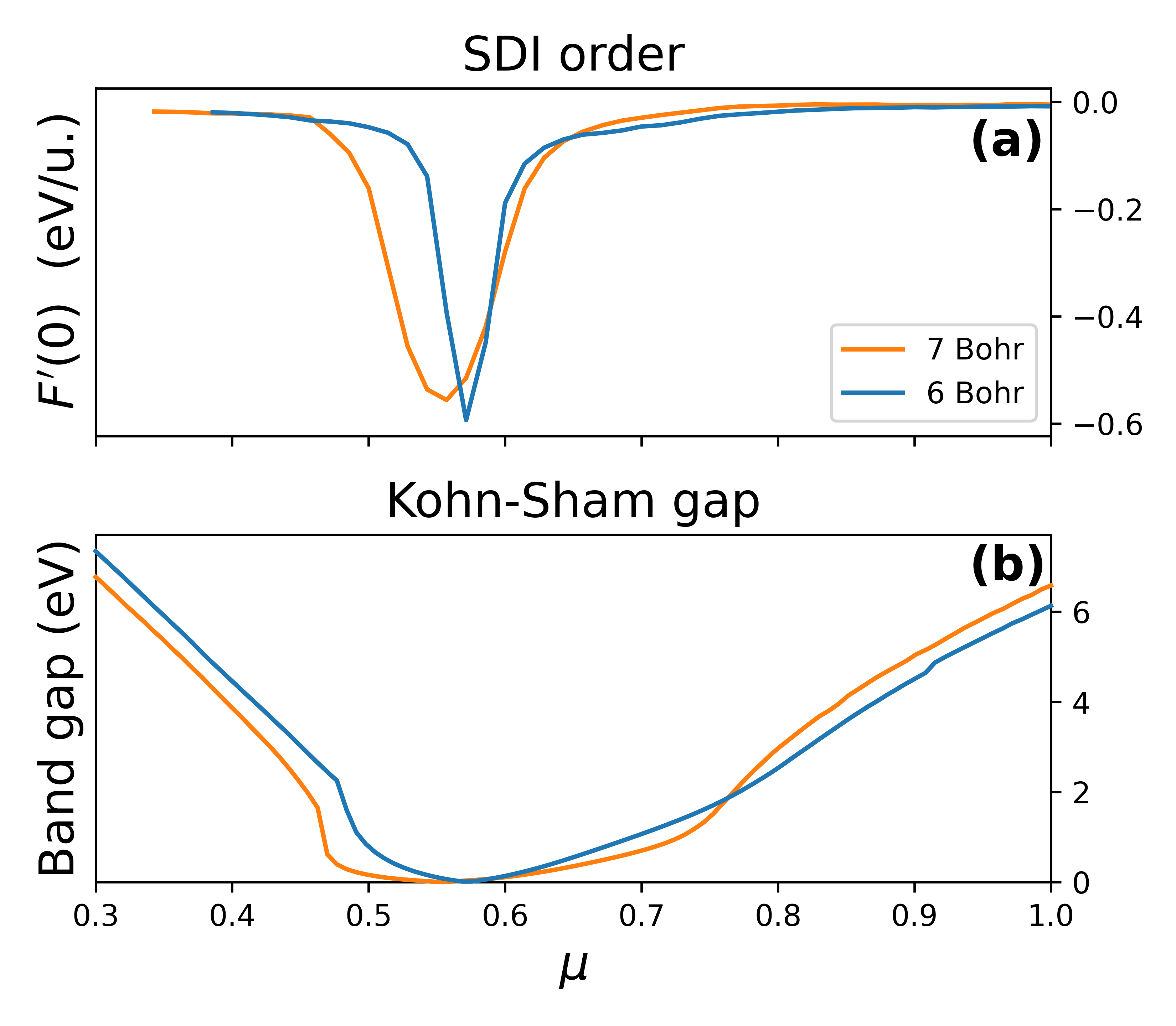}
        \caption{SDI phase obtained from DFT with explicit symmetry breaking via infinitesimal $\text{d}\rho$. (a) Order parameter for the SDI phase given by the first derivative of the free energy at $\rho=0$ for two sample geometries, $R_0 = 7$ Bohr and $R_0 = 6$ Bohr. (b) Kohn-Sham band gap as a function of $\mu$ (with $\text{d}\rho=0$) for the two geometries considered in (a). }
        \label{fig:sdi}
    \end{figure}

    \textit{Physical system.}---The electronic ground state of the NNC chain is a function of three free parameters: the bond length $R_0$, the charge ratio $\mu=Z_A/Z_B$, and the lattice distortion $\text{d}\rho$ (see Fig. \ref{fig:schematic}). Explicitly, we define the Hamiltonian (for $N$ cells) as
    \begin{equation}\label{eqn:nnc-hamiltonian}
        \hat H_{\text{NNC}} =\sum_{i=1}^{2N}\left(-\frac{1}{2}\nabla_i^2 + \hat V_L(r_i)\right) + \sum_{i<j}^{2N}\hat V_C(r_i,r_j),
    \end{equation}
    where 
    \begin{equation}\label{eqn:lattice-potential}
        \hat V_L(r_i) = -\frac{2}{\mu+1}\sum_{j=0}^{N-1}\left(\frac{\mu}{|r_i-R_j^A|} + \frac{1}{|r_i-R_j^B|}\right)
    \end{equation}
    is the periodic lattice potential,
    \begin{equation}\label{eqn:RA}
        R_j^A = 2jR_0,
    \end{equation}
    \begin{equation}\label{eqn:RB}
        R_j^B = (2j+1)R_0 + \text{d}\rho,
    \end{equation}
    and
    \begin{equation}\label{eqn:coulomb}
        \hat V_C(r_i,r_j) = \frac{1}{|r_i-r_j|}
    \end{equation}
    is the Coulomb interaction. In real materials, the nuclear charge numbers are always integers. Removing this restriction provides a convenient way to model a continuous staggered potential between two sublattices within DFT, as indicated by the relative widths of the potential wells in Fig. \ref{fig:schematic}. In fact, the NNC chain has a rich phase diagram owing to this continuity. Previously, it was well known that the 1D hydrogen chain (i.e., $\mu = 1$) is metallic at equilibrium and becomes insulating as it is stretched beyond $R_0\approx 1.71$ Bohr \cite{h-chain1,h-chain2}. It was later reported by Ref. \cite{nnc} that tuning $\mu$ enables charge transfer physics, causing the gap to vanish again at a critical point for fixed $R_0$. Based on this result, we have performed high-throughput, periodic, broken-symmetry DFT calculations (using \verb|VASP| \cite{vasp1, vasp2}) on the NNC chain, with the primary findings summarized in Figs. \ref{fig:dft-phase-diagram} and \ref{fig:sdi}. 

    \textit{DFT results.}---We give a general overview of the methodology here but defer the remainder of the computational details to the Supplemental Material (SM) \textsection I. In a real interacting system, the ground state can spontaneously break a symmetry of the Hamiltonian, which leads to Goldstone modes in the case of continuous symmetry \cite{altland-book}. This phenomenon can be simulated in DFT by infinitesimally perturbing the system, which enables the variational algorithm to stabilize a target ground state \cite{dft-sym-break}. 
    
    We first consider the case $\text{d}\rho = 0$  in Eqn. \ref{eqn:RB}. For the corresponding Hamiltonian in Eqn. \ref{eqn:nnc-hamiltonian}, an MI is naturally anticipated if the charge imbalance is relatively weak. Hence, we explicitly break the $SU(2)$ symmetry of the chain by introducing antiferromagnetic (AFM) ordering, with spin-up (-down) moments initialized on the A (B) sites. We then calculate the AFM order parameter, which we define as 
    \begin{equation}
        M_{\text{AB}} = \left[\int_{\mathcal S_{\text A}} dr - \int_{\mathcal S_{\text B}} dr \right] \ Q(r),
    \end{equation}
    where $\mathcal S_\alpha$ denotes the Wigner-Seitz sphere centered on site $\alpha$ and $Q(r) = n_{\uparrow}(r) - n_{\downarrow}(r)$ is the ground state spin density. When $\mu$ is close to one, we find that the MI phase with a finite $M_{\text{AB}}$ is favored. Decreasing $\mu$ increases the charge imbalance and suppresses the MI phase such that $M_{\text{AB}}$ vanishes discontinuously at some critical value, indicated by the black curve in Fig. \ref{fig:dft-phase-diagram}.
    We emphasize that this curve is not a phase transition in the true sense because a generic state is a superposition of many different $SU(2)$-broken configurations, requiring an enlarged unit cell. Moreover, a continuous symmetry cannot be spontaneously broken in 1D due to the famous Mermin–Wagner theorem \cite{altland-book}. While the exact location of the phase boundary is irrelevant to the current work, the fact that an explicit symmetry breaking field lowers the energy of the AFM state relative to the nonmagnetic (NM) state is consistent with a 1D MI. 

    As the ground state configuration changes according to DFT, we expect a gapless point to emerge in the effective Kohn-Sham system describing the NNC chain. To precisely determine this point, we first use \verb|Wannier90| \cite{w90} to fit the DFT band structures to tight-binding models. We then use \verb|Z2Pack| \cite{z2pack} to calculate the total Berry phase in the occupied subspace,
    \begin{equation}\label{eqn:berry}
        \gamma = \sum_{\sigma=\uparrow,\downarrow}\int dk\ \braket{u^{\sigma}_{k}|i\partial_k|u^{\sigma}_{k}}.
    \end{equation}
    In the above expression, $\ket{u_k^{\sigma}}$ denotes the lowest Bloch state at momentum $k$ in the spin-$\sigma$ sector of the Kohn-Sham eigenspectrum. Because Eqn. \ref{eqn:nnc-hamiltonian} respects site-centered inversion symmetry when $\text{d}\rho=0$, $\gamma$ is guaranteed to be quantized as originally proven by Zak \cite{zak}. Two chains with different Berry phases cannot be deformed into each other without closing the gap \cite{berry-phases-vanderbilt}, and so this technique of identifying the phase boundary is sometimes called the method of topological transitions (MTT) \cite{mtt}. Alternatively, one can interpret this transition as a fine-tuning of the \textit{effective} staggered potential in the tight-binding model, which we describe in the SM \textsection II. Similarly, the AFM transition can be understood in the context of an analytical mean-field model, which we have also included in SM \textsection III. 

    \textit{Inversion symmetry and SDI.}---While the inversion symmetry guarantees the topological number and hence the validity of the MTT, we should consider the possibility of broken inversion symmetry. If the symmetry-respecting ground state exhibits an instability with respect to an infinitesimal $\text{d}\rho$, then the true ground state must spontaneously break inversion symmetry \cite{sachdev,quantum_theory_of_e_liquid}, which is precisely the SDI phase. According to Landau theory, the free energy $F$ is an even functional of the SDI order parameter $\mathcal D$, and it follows that $F'(\rho)\propto \mathcal D\cdot \mathcal D'(\rho)$ up to first order. Thus, $F'$ can be used to probe the SDI phase of the NNC chain as shown in Fig. \ref{fig:sdi}(a). To demonstrate the general trend, we consider two different lattice geometries, $R_0 = 6$ and 7 Bohr. We observe an appreciable SDI regime at larger $\mu$ for smaller $R_0$. The magnitude of the order parameter reaches a maximum near the gapless point, shown in Fig. \ref{fig:sdi}(b), which is consistent with the interpretation of $\mu$ as a pumping parameter as explained in SM \textsection II.

    \textit{Model comparison.}---We provide a benchmark for our DFT calculations by comparing them to an exactly solvable model. As the essential physics originates from the competition between the staggered potential and the Coulomb interaction, the IHM provides a faithful and feasible benchmark. The IHM (with one orbital per site) is defined as 
    \begin{equation}\label{eqn:ihm}
\hat H_{\text{IHM}} = \hat H_t + \hat H_U + \hat H_{\Delta},
\end{equation}
where 
\begin{equation}
    \hat H_t = -t\sum_{j,\sigma}\left(\hat c_{A\sigma,j}^{\dagger}\hat c_{B\sigma,j} + \hat c_{B\sigma,j}^{\dagger}\hat c_{A\sigma,j+1} + \text{h.c.}\right)
\end{equation}
represents nearest-neighbor hopping,
\begin{equation}
    \hat H_U = U\sum_{j,\alpha}\hat n_{\alpha\uparrow,j}\hat n_{\alpha\downarrow,j} 
\end{equation}
is the Hubbard interaction, and
\begin{equation}
    \hat H_{\Delta} = \Delta\sum_{j, \sigma}\left(\hat n_{A\sigma,j} -\hat n_{B\sigma,j}\right)
\end{equation}
represents the staggered potential. In the above sums, $j$ denotes the unit cell, $\sigma$ denotes spin ($\uparrow$ or $\downarrow$), and $\alpha$ denotes the sublattice (A or B). 

Next, we detail the mapping between the NNC chain in Eqn. \ref{eqn:nnc-hamiltonian} and the IHM in Eqn. \ref{eqn:ihm}, and 
we argue that the IHM captures the essential physics of the NNC chain up to leading order in the interaction. As the bond length increases, hopping between sites is suppressed such that the interaction strength $U/t$ is large, approaching infinity in the atomic limit. In the DFT framework, this corresponds to an exponential decay of the density overlap between local electron orbits at each nucleus. We therefore write the hopping as an exponentially decaying function of $R_0$, i.e., $t=c_1e^{-R_0/c_2}$, based on hydrogen-like one-electron orbitals. Then, to recover the behavior in the atomic limit, we set $U$ to be constant. Lastly, we let $\mu$ be the ratio of the one-electron densities on the two sublattices as determined by the Bloch Hamiltonian for a free chain (with $U=0$). A straightforward derivation yields
\begin{equation}
    \mu = \frac{\frac{1}{2} - \eta(t,\Delta)}{\frac{1}{2} + \eta(t,\Delta)},
\end{equation}
where
\begin{equation}
    \eta(t,\Delta) = \frac{\Delta }{\pi\sqrt{4t^2 + \Delta^2}}K\left(\frac{4t^2}{4t^2 + \Delta^2}\right)
\end{equation}
and
\begin{equation}
    K(m) = \int_0^{\pi/2} \frac{d\theta}{\sqrt{1 - m \sin^2\theta}}
\end{equation}
is the complete elliptic integral of the first kind. In practice, $\Delta$ can be determined as a function of $\mu$ (up to an appropriate scale factor) using Newton's method. One can easily verify that this function is monotonically decreasing as expected (see Fig. S2).

\begin{figure}
    \centering
    \includegraphics[width=\linewidth]{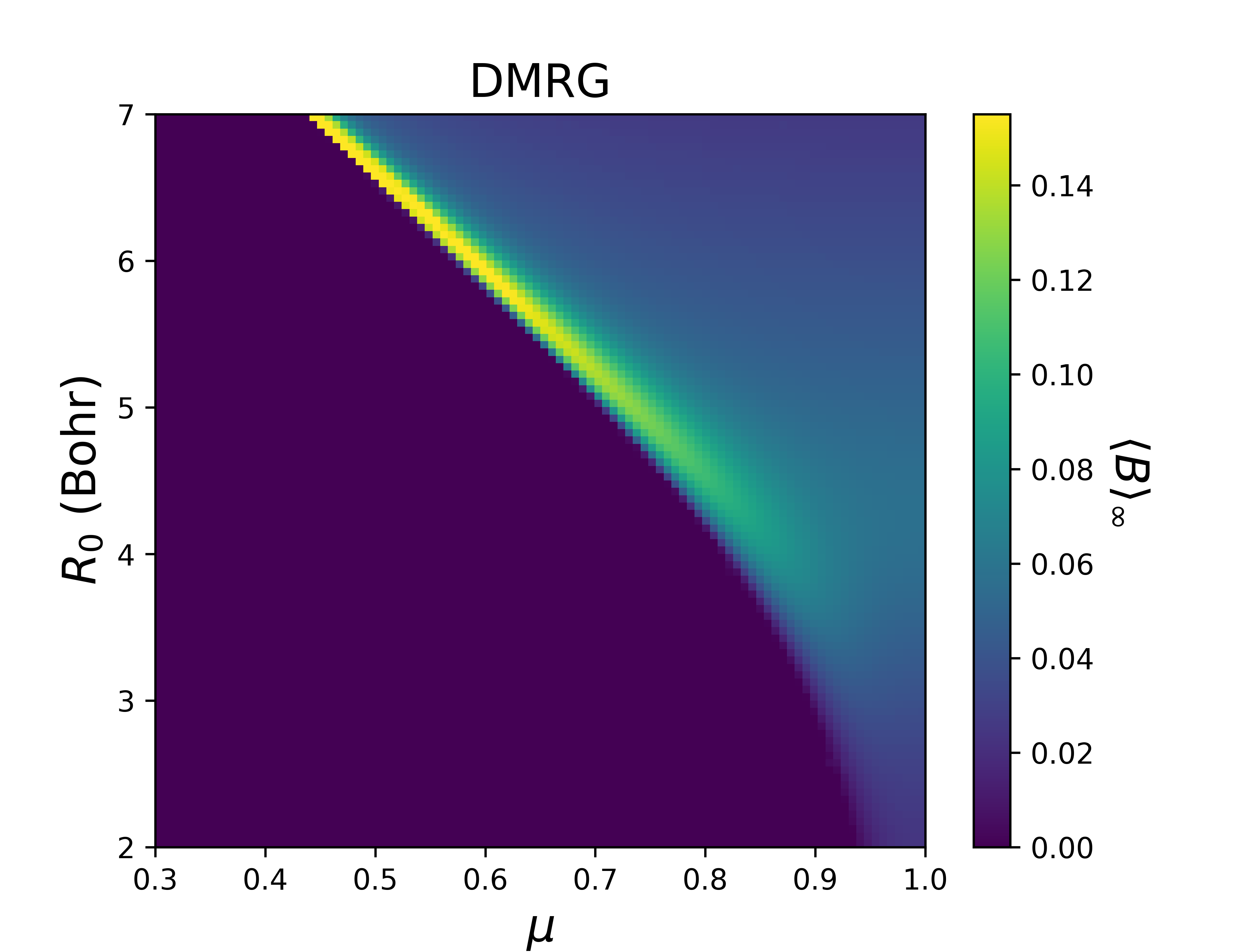}
    \caption{Phase diagram of the IHM obtained via finite DMRG with open boundary conditions and realistic model parameters (see main text). The color bar represents the expectation value of the bond order operator, extrapolated to an infinite chain.}
    \label{fig:dmrg}
\end{figure}

Having established the mapping, we are ready to  compare the phase diagram of the IHM to the NNC chain. To that end, we calculate the SDI order parameter by taking the expectation value of the bond order operator \cite{ihm-bosonization}, defined as
\begin{equation}
B = \sum_{j,\sigma}\left(\hat c^{\dagger}_{A\sigma,j}\hat c_{B\sigma,j}-\hat c^{\dagger}_{B\sigma,j}\hat c_{A\sigma,j+1} + \text{h.c.}\right).
\end{equation}
The many-body ground state is determined using DMRG, which is considered to be highly reliable in 1D \cite{dmrg}. Broken symmetry is achieved by performing DMRG calculations in \verb|TeNPy| \cite{tenpy} on a finite chain with open boundary conditions and extrapolating to an infinite chain. We note that similar work has been reported in the literature \cite{ihm-dmrg}, and the following results are qualitatively consistent. To quantitatively compare to the DFT phase diagram, the model parameters are chosen as $U=1$, $c_1 = 2.5$, and $c_2 = 2.2$. The remaining computational details are included in the Supplemental Material \textsection I. The resulting phase diagram is presented in Fig. \ref{fig:dmrg}. The SDI regime is clearly visible and follows the same general trend as the DFT result. For example, consider the behavior at $R_0 = 7$ Bohr. For $\mu$ less than $\around$0.44, the bond ordering is completely suppressed because both $U/t$ and $\Delta$ are large; for $\mu$ greater than $\around$0.44, the ground state spontaneously breaks inversion symmetry. This transition occurs at larger $\mu$ for smaller $R_0$. 

To give additional context to our calculations, we briefly mention that the IHM hosts three distinct phases for a fixed $\Delta$ as reported in Ref. \cite{ihm-bosonization}. At $U=0$, the system is a BI, and it undergoes an Ising transition to an SDI phase at a critical interaction, where the charge gap closes and reopens. Further increasing $U$ results in a Kosterlitz-Thouless (KT) transition to an MI phase, where the spin gap remains closed. Additionally, Ref. \cite{ihm-mtt} reported the same phase diagram using the MTT. In the BI (MI) phase, the charge and spin Berry phases are both zero ($\pi$). In the SDI phase, the charge Berry phase is $\pi$, while the spin Berry phase is zero. 

Since the SDI is well understood to be the intermediate phase between the BI and MI in the IHM, we did not pursue these two phases explicitly in the current DMRG work. In DFT, there is no straightforward way to simulate spin-charge separation. Instead, we argue that the AFM (NM) ground state of the NNC chain is analogous to the MI (BI) phase in the IHM. The key difference is that the DFT phase diagram appears to host simultaneous AFM and SDI ordering around the gapless point. We identify two possible explanations for this. The first is that it is a consequence of the AFM setting in DFT as previously discussed. It is possible that different $SU(2)$-broken states will reverse the order of the black and red curves in Fig. \ref{fig:dft-phase-diagram} such that magnetic ordering only occurs when the SDI phase is totally suppressed. This would require a larger supercell and could be explored in future DFT work. The second explanation is that the KT transition is not accessible to DFT because the Kohn-Sham band gap is a mean-field-like description of the spectral gap for the real NNC chain. In the SDI regime, it is not possible to quantize the single-particle Berry phase because the inversion symmetry is broken. However, this does not mean that evidence of the transition is totally absent in DFT. In fact, we observe that the band gap is discontinuous along the black curve in Fig. \ref{fig:dft-phase-diagram}. According to a naive mean-field model, this discontinuity represents a first-order Landau phase transition, as we discuss in SM \textsection III. In this sense, DFT is robust because it provides evidence for all three phases present in the IHM while incorporating the full Coulomb interaction (with the only approximation being the exchange-correlation functional).

\textit{Discussion.}---We now discuss how the preceding analysis provides insight into the capability of DFT to capture interaction in real materials. We first note that the NNC method has previously been used in combination with spin symmetry breaking by Ref. \cite{nnc} to simulate doping in the quasi-1D cuprate chain Ba$_2$CuO$_3$. While such applications with artificial NNC are likely limited to a few types of materials, we emphasize that the NNC chain studied here is simply a benchmark for symmetry-broken DFT in general. Similar to our SDI analysis, lattice degrees of freedom have been shown to play a crucial role in stabilizing correlated phases of real materials, like the various stripe and magnetic phases of the cuprate YBa$_2$Cu$_3$O$_7$ \cite{ybco}. Additionally, DFT with broken translational symmetry has been used to accurately predict the charge density wave ground state of the superconducting kagome metal family XV$_3$Sb$_5$ (X = K, Rb, or Cs) \cite{cdw}. Our current study provides essential context to these types of applications. In particular, lattice distortions effectively add source fields to the Landau free energy functional describing the material. We have demonstrated that, up to first order, the spontaneous symmetry breaking induced by these fields correctly reflects the exact phase diagram, even in the strongly interacting regime (i.e., $R_0 = 7$ Bohr). Moreover, the ground state total energy of DFT is formally exact and can be highly accurate, especially when advanced functionals are used. Hence, our results formally validate the methodology as a reliable tool for predicting real material phases.


\textit{Concluding remarks.}---In closing, we have determined the first-principles phase diagram of an interacting ionic chain using DFT with broken symmetry and non-integer nuclear charge numbers. Our methodology is benchmarked against an independent DMRG calculation via a simple but robust mapping from the NNC chain to the IHM.
The results demonstrate that DFT, despite its mean-field-like construction, is capable of directly accessing symmetry-broken phases of real materials in a way that is qualitatively consistent with exact many-body approaches.

\textit{Acknowledgement.}---J.K. acknowledges the support from the National Science Foundation (NSF) Graduate Research Fellowship Program under Grant No. 2139911. J.S. acknowledges the support from the NSF under Grant No. DMR-2042618. Any opinions, findings, and conclusions or recommendations expressed in this material are those of the author(s) and do not necessarily reflect the views of the NSF. Our work has used Purdue Anvil CPUs through allocation DMR190076 from the Advanced Cyberinfrastructure Coordination Ecosystem, as well as the National Energy Research Scientific Computing Center.

\raggedright
\bibliography{refs}



\end{document}


\title{Supplemental Material: First-principles phase diagram of an interacting ionic chain}
\author{Jamin Kidd}
\author{Ruiqi Zhang}
\author{Shao-Kai Jian}
\author{Jianwei Sun}
    \email{jsun@tulane.edu}
\affiliation{\mbox{Department of Physics and Engineering Physics, Tulane University, New Orleans, Louisiana 70118, USA}}
\date{\today}

{
\let\clearpage\relax
\maketitle
}
\section{Computational details}\label{sec:methods}
\textit{DFT.}---The phase diagram in Fig. 2 of the main text is obtained by performing high-throughput periodic DFT calculations on a $100\times100$ grid of chain parameters in \verb|VASP| \cite{vasp1,vasp2}. For each configuration, the lattice is embedded into a supercell with large vacuum layers of 56.7 Bohr in the two directions orthogonal to the chain. To simulate non-integer nuclear charge numbers at each site, charges are adjusted in the projector-augmented-wave pseudopotentials \cite{paw1,paw2}. All calculations use the r$^2$SCAN meta-GGA functional \cite{r2scan} with a $40\times1\times1$ $k$-point mesh, a basis set cutoff energy of $600$ eV, and a convergence criterion of $10^{-7}$ eV \footnote{We have also repeated the DFT calculations using the PBE GGA functional \cite{pbe} and found no qualitative difference in the phase diagram}. Two-band tight binding models are obtained for each spin sector by projecting the DFT Bloch states onto trial $s$-like orbitals at each site and fitting to the band dispersion using the \verb|Wannier90| code \cite{w90}. The quality of the fitting is ensured by checking that the DFT and tight binding dispersions match point-wise, up to a tolerance equal to
1\% of the bandwidth. A sample of such fittings is shown in Fig. S1. The Berry phase for each chain is computed using the method of hybrid Wannier charge centers as implemented in \verb|Z2Pack| \cite{z2pack}. For the calculations with broken inversion symmetry, the lattice is dimerized under linear response in the free energy and then extrapolated back to the original chain (i.e., $\text{d}\rho = 0$). The position of nucleus B is shifted inside the cell by $\text{d}\rho = 2\varepsilon R_0$. Three such geometries are considered: $\varepsilon = 10^{-3}$, $5\times10^{-4}$, and $10^{-4}$. For the fitting, the minimum coefficient of determination across all chain configurations is found to be 0.903, indicating high quality.

\textit{DMRG.}---Fig. 4 of the main text is obtained by performing finite DMRG calculations with open boundary conditions in \verb|TeNPy| \cite{tenpy} on the same $(R_0, \mu)$ grid used for the DFT calculations. Each chain is mapped to the IHM following the discussion in the main text. The staggered potential $\Delta$ is fine-tuned by an overall scale factor of 10 such that the SDI phase is visible in the considered range of $Z_A/Z_B$ values (see Eqn. 12 of the main text and Fig. \ref{fig:delta} below). The matrix product state is truncated such that all Schmidt values less than $10^{-6}$ are discarded and no more than 800 total Schmidt values are kept. The ground state is obtained variationally, with a convergence criterion of $10^{-5}$ energy units. To obtain the SDI order parameter in the limit of an infinite chain, we calculate $\braket{\hat B}$ for different chain lengths $L$ and fit to the power-law function $\braket{\hat B} = \kappa_1L^{-\kappa_2} + \braket{\hat B}_{\infty}$ with $0\leq\kappa_2\leq 1$. For $L\in\{10,12,14,16\}$, we find that an accurate fitting is already achieved, with the largest residual sum of squares across the whole parameter space being $3.34\times10^{-5}$ squared energy units.
\begin{figure}
    \centering
    \includegraphics[width=\linewidth]{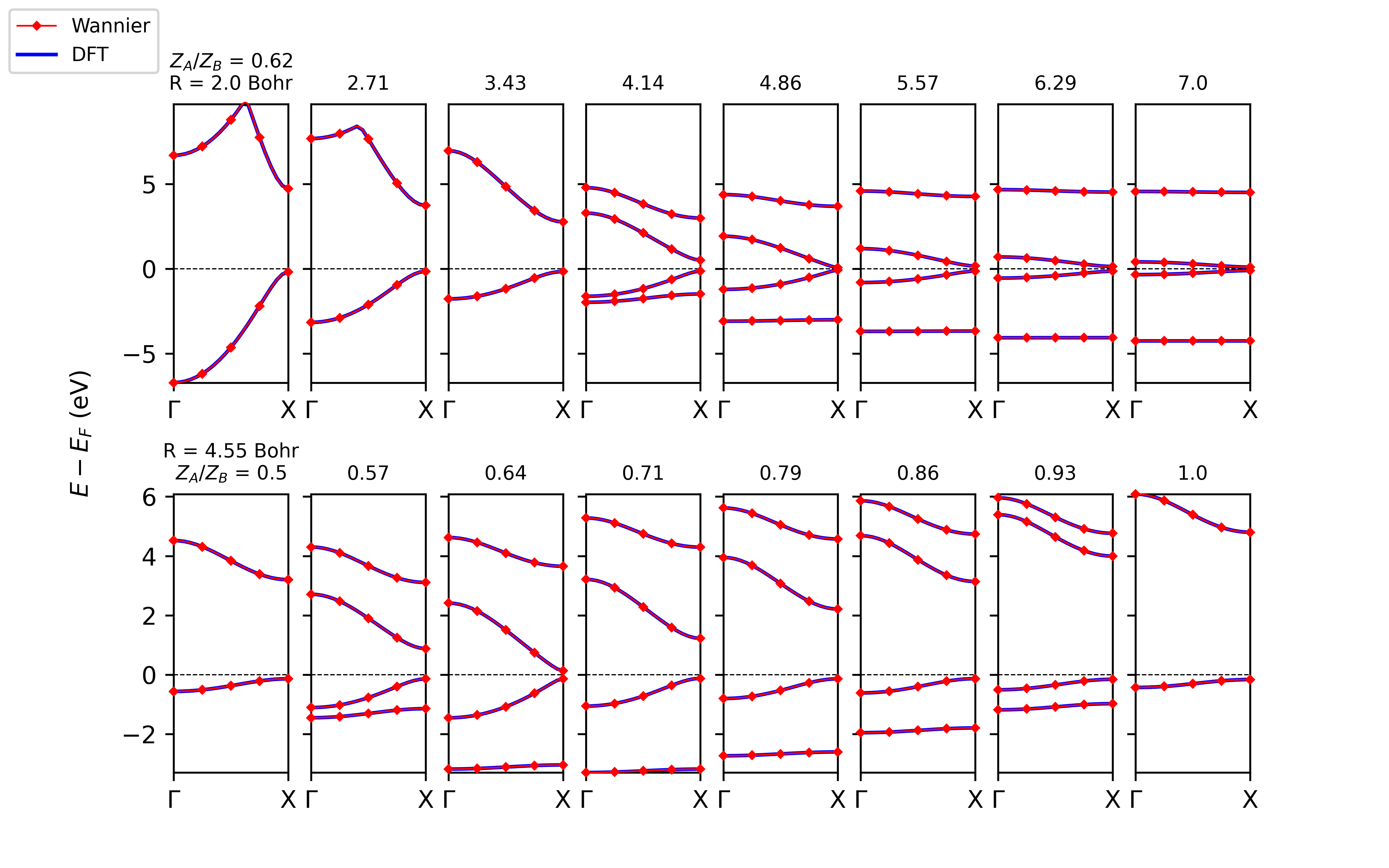}
    \caption{A selection of DFT-obtained band structures (blue) with Wannier fittings (red) for the NNC chain. (Top) $Z_A/Z_B = 0.62$ and $R_0$ varies from 2.0 Bohr to 7.0 Bohr. (Bottom) $R_0 = 4.55$ Bohr and $Z_A/Z_B$ varies from 0.5 to 1.0. Since $\text{d}\rho=0$, the chain respects site-centered inversion symmetry as discussed in the main text. The gapless point in the phase diagram corresponds to a direct closing of the Kohn-Sham band gap at the Brillouin zone boundary X.}
    \label{fig:tb}
\end{figure}
\begin{figure}
    \centering
    \includegraphics[width=0.6\linewidth]{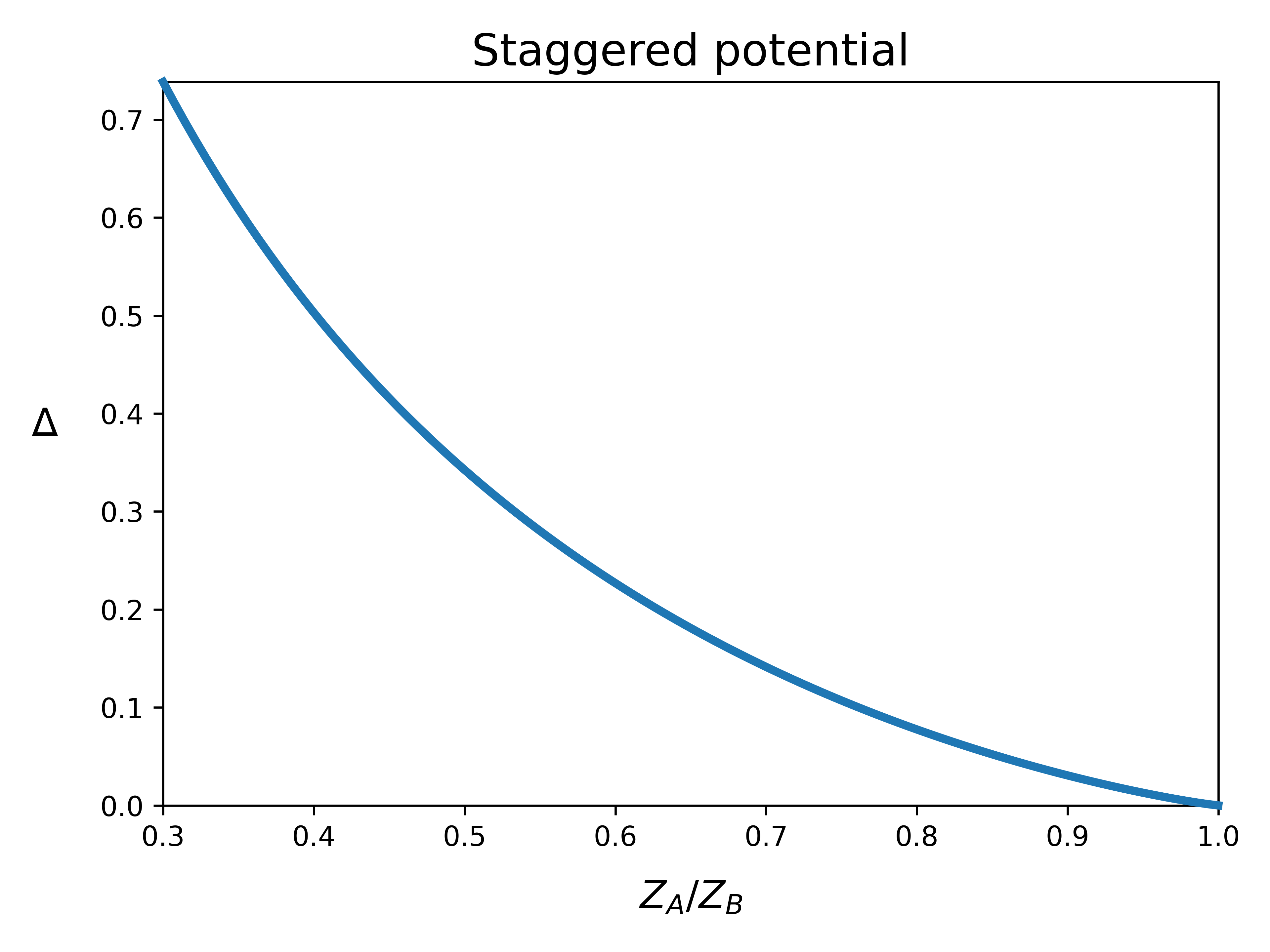}
    \caption{Staggered potential $\Delta$ as a function of $Z_A/Z_B$, obtained by applying Newton root-finding to Eqn. 12.}
    \label{fig:delta}
\end{figure}

\section{Tight-binding analysis}
\begin{table}[htp]
    \centering
    \begin{tabular}{c|c|c|c|c}
        \hline
        \hline & & & & \\[-2ex]
        $\mu$ & $\Delta^{(0)}_\uparrow$ & $t^{(0)}_\uparrow$ & $\Delta^{(1)}_\uparrow$ & $t^{(1)}_\uparrow$\\[1ex]
        \hline
        1 & -3.3140  & 0.1642 & 0.0122 & 0.0053 \\
        \hline
        0.5545 & -0.0005 & 0.1641  & -0.0003 & 0.0049 \\
        \hline
        0.3 & 3.3931 & 0.1181  & -0.0036 & 0.0037 \\
        \hline
        \hline
    \end{tabular}
    \quad \quad \quad
    \begin{tabular}{c|c|c|c|c}
        \hline
        \hline & & & & \\[-2ex]
        $\mu$ & $\Delta^{(0)}_\downarrow$ & $t^{(0)}_\downarrow$ & $\Delta^{(1)}_\downarrow$ & $t^{(1)}_\downarrow$\\[1ex]
        \hline
        1 & 3.3140  & 0.1642&  -0.0122 & -0.0014 \\
        \hline
        0.5545 & 3.3022 & 0.1007  & -0.0062 & 0.0028 \\
        \hline
        0.3 & 3.3931 & 0.1181 &  -0.0036 & 0.0037 \\
        \hline
        \hline
    \end{tabular}
    \caption{Leading order tight-binding couplings at $R_0 = 7$ Bohr for select values of $\mu$ in the spin-up (left) and spin-down (right) sectors. $\Delta^{(0)}_\tau$ represents a staggered potential, while $t^{(0)}_\tau$, $\Delta^{(0)}_\tau$, and $t^{(1)}_\tau$ represent nearest, next-nearest, and next-next-nearest neighbor hopping, respectively.} 
    \label{tab:couplings}
\end{table}

Here, we provide a more detailed analysis of the tight binding models obtained from our inversion-symmetric (i.e., $\text{d}\rho=0$) DFT calculations that are used to determine the phase diagram in Fig. 1 of the main text. We first review the generic form of a spinful four-band tight-binding model without spin-orbit coupling. In momentum space, the Hamiltonian is 
\begin{align}
    H_k & =  \mathbbm{1} \otimes  \left(\mathbf d^\tau_k \cdot \vec\sigma\right)
\end{align}
where $\mathbbm{1}$ is the $2\times2$ identity matrix, $\tau$ denotes spin ($\uparrow$ or $\downarrow$), and $\vec\sigma = \left(\sigma^0,\sigma^x, \sigma^y, \sigma^z\right)$ is the Pauli vector. Without loss of generality, we assume that $H_k$ is traceless so that the $\sigma^0$-component of $\mathbf d_k^\tau$ is zero \footnote{This assumption is allowed because the $\sigma^0$-component affects neither the presence nor location (in $k$-space) of the direct gap closing.}. In our convention,
\begin{equation}\label{eqn:d-vec}
    \mathbf{d}^\tau_k = \renewcommand\arraystretch{2.4}\begin{pmatrix}
        0\\
        -t_\tau^{(0)}\\
        0\\
        \Delta_\tau^{(0)}
    \end{pmatrix} + \ \sum_{n=1}^{N_k-1} \ \begin{pmatrix}
        0 \\
         -  \left[t_\tau^{(n)} + t_\tau^{(n-1)}\right] \ \cos \ (nk)\\
        \left[t_\tau^{(n)} - t_\tau^{(n-1)}\right] \ \sin \ (nk)\\
        2 \Delta_\tau^{(n)}\ \cos \ (nk)
    \end{pmatrix}
\end{equation}
where
\begin{equation}
    \Delta_\tau^{(n)}=\frac{1}{2}\left(\Delta_{A,\tau}^{(n)} - \Delta_{B,\tau}^{(n)}\right),
\end{equation}
$\Delta_{\alpha,\tau}^{(n)}$ denotes hopping within sublattice $\alpha$ (= A or B), $t_\tau^{(n)}$ denotes hopping from B to A, and $n$ is the cell displacement. 

For larger $R_0$, extended hopping terms tend to become perturbative. To demonstrate this point, we consider the example of the chain with $R_0 = 7$ Bohr. We have extracted the couplings as functions of $\mu$ by applying inverse discrete trigonometric transforms to Eqn. \ref{eqn:d-vec} on a finite mesh of $N_k = 40$ grid points. The vectors $\mathbf d_k^\tau$ are calculated directly from the DFT models using the \verb|TBmodels| interface \cite{tbmodels}. The relevant parameters obtained for each spin sector are given in Table \ref{tab:couplings}. We find that couplings beyond nearest-neighbor hopping are strongly suppressed by the large bond length. Hence, the Kohn-Sham spectrum is effectively described by a Rice-Mele model, i.e.,
\begin{equation}
    \mathbf{d}^\tau(k) = 
    \renewcommand\arraystretch{2}\begin{pmatrix}
        0\\
        -t^{(0)}_\tau\left(1 + \cos k \right)\\
        -t^{(0)}_\tau\sin k\\
        \Delta_\tau^{(0)}
    \end{pmatrix}
\end{equation}
yielding the band structure
\begin{equation}
    E_\tau(k) = \pm\sqrt{2\left[t^{(0)}_\tau\right]^2(1+\cos k) + \left[\Delta^{(0)}_\tau\right]^2}
\end{equation}
At $k=\pi$, there is a direct band gap
\begin{equation}\label{eqn:gap}
    \Delta = \min\left(2\left|\Delta^{(0)}_\uparrow\right|,\ 2\left|\Delta^{(0)}_\downarrow\right|\right).
\end{equation}

\begin{figure}[h]
    \centering
    \includegraphics[width=0.7\linewidth]{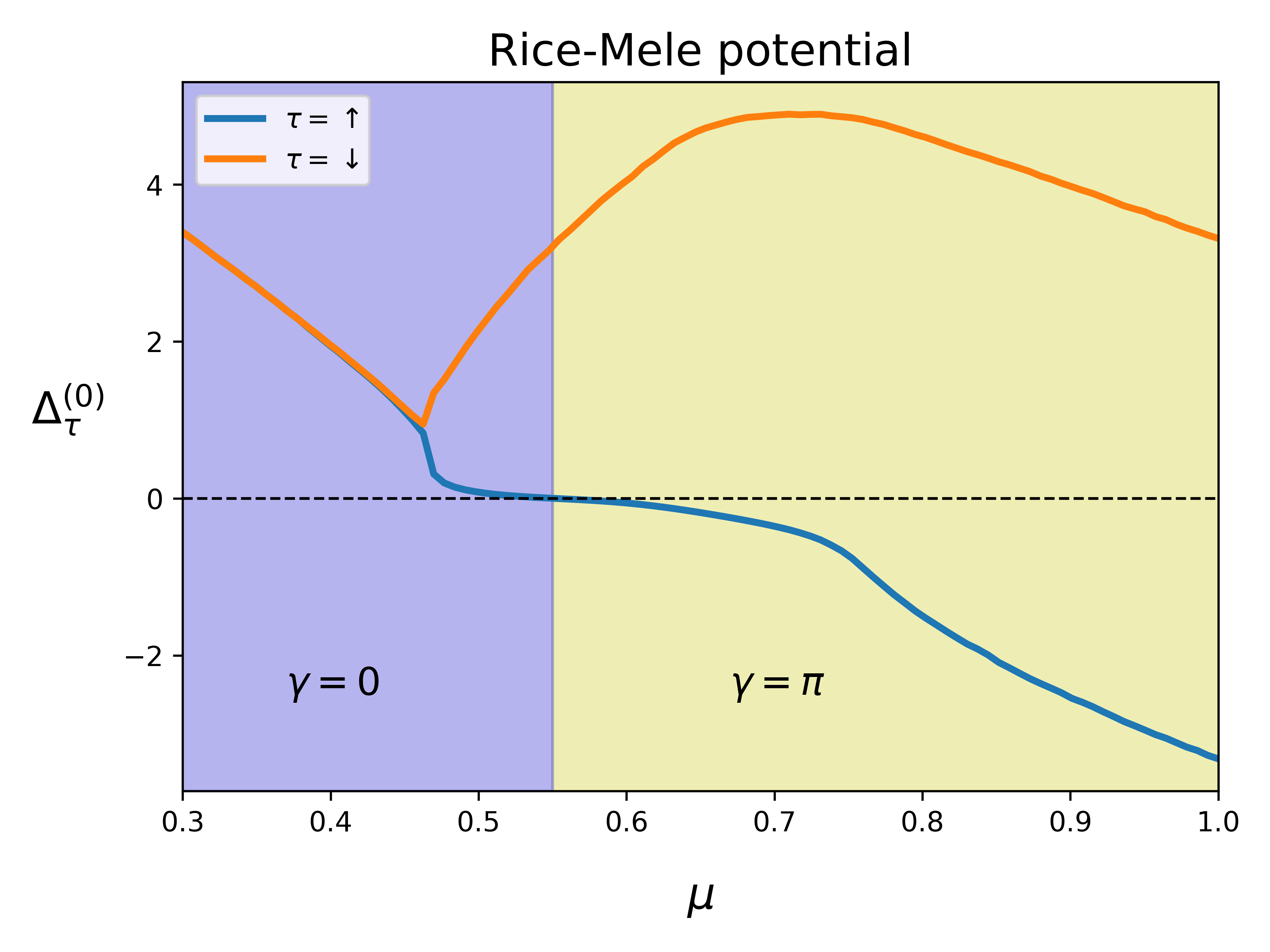}
    \caption{Effective staggered potential obtained from DFT as a function of $\mu$. The Mott (band) insulator phase is characterized by negative (positive) staggering in the spin-up sector. In the MTT, the sign maps to the quantized Berry phase $\gamma$.
    The band gap closes at the phase transition near $\mu = 0.55$. Near $\mu = 0.46$, staggering is no longer spin-polarized because the phase is nonmagnetic.}
    \label{fig:rice-mele}
\end{figure}

The relevance of the above result to the DFT phase diagram can be seen via the following thought experiment. Consider the large-$R_0$ limit of an NNC chain in which the electrons are non-interacting. At $\mu = 1$, the chain is metallic. For $\mu < 1$, the chain is a band insulator in which electrons doubly occupy sublattice B. Meanwhile, DFT predicts an insulator phase with non-zero $\Delta$ at $\mu=1$ because the tight-binding model implicitly includes Coulomb interaction under the Kohn-Shame scheme. Because $SU(2)$ symmetry is broken in this phase, we expect that $\Delta^{(0)}_\uparrow\neq\Delta^{(0)}_\downarrow$. By tuning $\mu$, the effective potential in one of the spin sectors decreases in magnitude and eventually reaches a critical
point where it switches signs. This point is precisely the gapless point according to Eqn. \ref{eqn:gap}. Hence, $\mu$ can be viewed as a pumping parameter in the context of the MTT. By further decreasing $\mu$, magnetic ordering is suppressed, and the trivial band insulator phase of the non-interacting chain is recovered. To illustrate this trend, we have plotted $\Delta^{(0)}_\tau$ as a function of $\mu$ in Fig. \ref{fig:rice-mele}.

\clearpage
\section{Analytical mean-field analysis}\label{sec:meanfield}
Starting from $\hat H_{\text{IHM}}$ in Eqn. 8 of the main text, we introduce an auxilliary field $n_{\alpha}$ such that 
\begin{equation}
    \hat n _{\alpha\uparrow,j}\hat n _{\alpha\downarrow,j} = n_{\alpha}^2 -n_{\alpha}(\hat n _{\alpha\uparrow,j}-\hat n _{\alpha\downarrow,j})+\delta_{\alpha},
\end{equation}
where 
\begin{equation}
    \delta_{\alpha} = (\hat n _{\alpha\uparrow,j}-n_{\alpha})(\hat n _{\alpha\downarrow,j}+n_{\alpha}).
\end{equation}
Letting $n_B=-n_A=-n$, the Hubbard term in $\hat H_{\text{IHM}}$ becomes
\begin{equation}
    \hat H_U\approx 2NUn^2 - Un\sum_{j}\left(\hat n_{A\uparrow,j}-\hat n_{B\uparrow,j} - \hat n_{A\downarrow,j} + \hat n_{B\downarrow,j} \right),
\end{equation}
where $N$ is the number of cells. The band structure is found to be
\begin{equation}\label{eqn:mean-field-bands}
    [E(k)]^2 = (Un\pm\Delta)^2 + 2t^2(1+\cos{k}),
\end{equation}
and so there is a direct gap closing at $k=\pi$ when $Un=\Delta$. The total energy of occupied bands per cell is 
\begin{equation}\label{eqn:energy-density}
    \varepsilon(n)=2Un^2-\frac{2}{\pi}\sum_{s=\pm1}\ \sqrt{(Un+s\Delta)^2+4t^2}\ E\left(\frac{4t^2}{(Un+s\Delta)^2+4t^2}\right)
\end{equation}
where 
\begin{equation}
    E(m)=\int_{0}^{\pi/2}d\theta \ \sqrt{1-m\sin^2\theta}
\end{equation}
is the complete elliptic integral of the second kind. Notice that $\varepsilon$ is an even function of $n$ and can therefore be expanded as 

\begin{figure}
    \centering
    \includegraphics[width=0.7\textwidth]{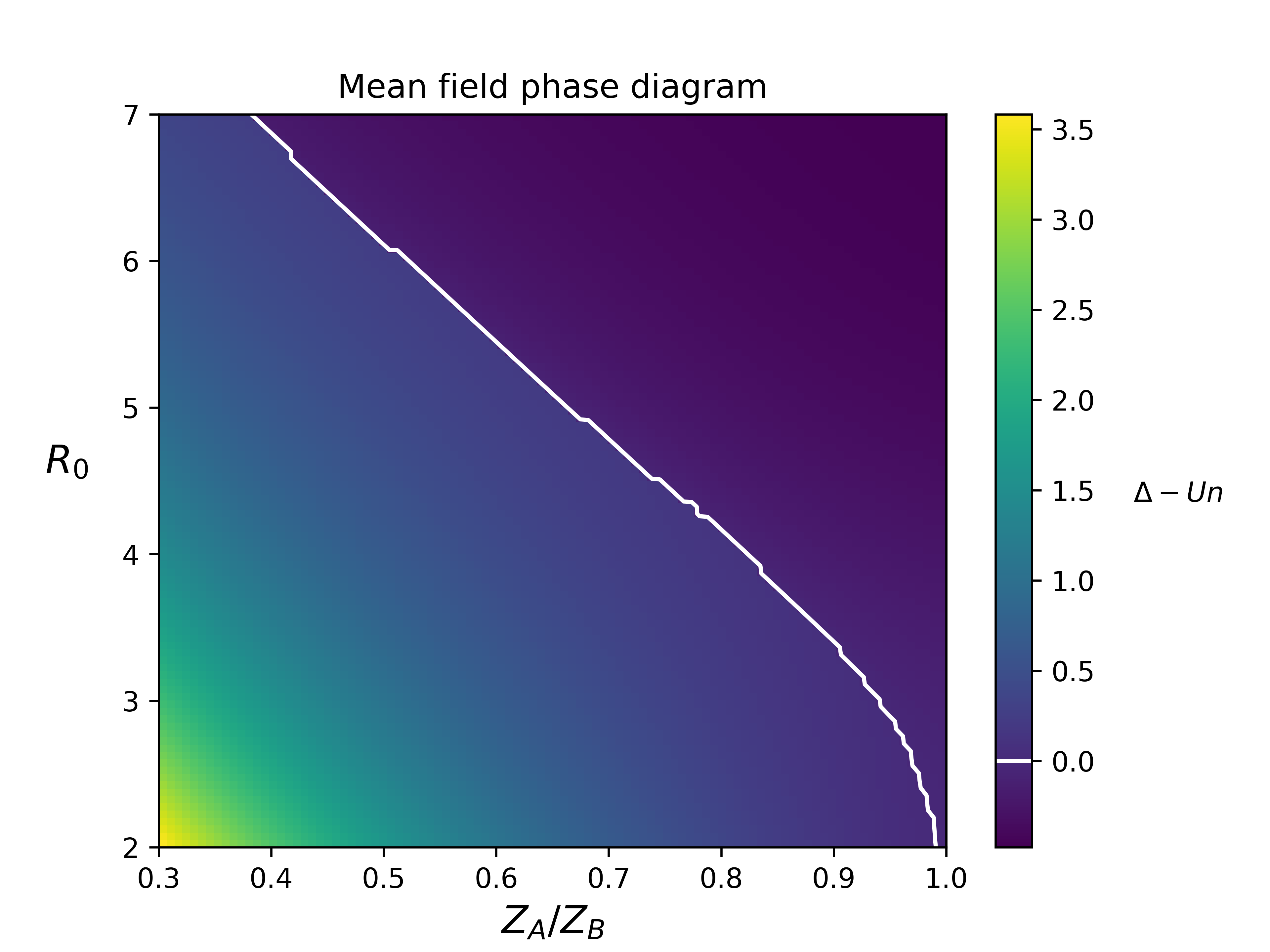}
    \caption{Phase diagram of the analytical mean-field model determined by self-consistently minimizing the energy density in Eqn. \ref{eqn:energy-density}. The quantity $\Delta-Un$ characterizes the gap. The boundary between positive and negative values is indicated by the white contour and represents a first-order phase transition. Near $Z_A/Z_B = 1$, the ground state favors AFM ordering (i.e., $n>0$). As $Z_A/Z_B$ is decreased for fixed $R_0$, the NM ground state (i.e., $n = 0$) becomes favored. This transition occurs at larger $Z_A/Z_B$ for smaller $R_0$, consistent with the DFT result in the main text.}
    \label{fig:meanfield}
\end{figure}

\begin{equation}
    \varepsilon(n) = c_0 + c_2n^2 + c_4n^4 + \mathcal{O}(n^6).
\end{equation}
We find the coefficients to be
\begin{equation}
    c_0 = -\frac{4}{\pi}\sqrt{\Delta^2+4t^2}E\left(\frac{4t^2}{\Delta^2+4t^2}\right),
\end{equation}
\begin{equation}
    c_2 = 2U-\frac{2}{\pi}U^2\left(\Delta^2+4t^2\right)^{-1/2}f(\Delta,t),
\end{equation}
and
\begin{equation}\label{eqn:c4}
    c_4 = -\frac{1}{6\pi}U^4\left(\Delta^2+4t^2\right)^{-5/2}\left[20t^2+\frac{16t^4}{\Delta^2}+(2\Delta^2-4t^2)f(\Delta,t)\right],
\end{equation}
where
\begin{equation}
    f(\Delta,t) \vcentcolon= K\left(\frac{4t^2}{\Delta^2+4t^2}\right)-E\left(\frac{4t^2}{\Delta^2+4t^2}\right),
\end{equation}
and $K$ is the complete elliptic integral of the first kind (Eqn. 14 of the main text). One can easily verify that the term in square brackets in Eqn. \ref{eqn:c4} is positive, and hence $c_4$ is negative. Therefore, $\varepsilon$ has three local minima corresponding to $n=0$ and $n=\pm n^*$ for some non-zero $n^*$. By tuning $c_2$, the global minimum jumps discontinuously, indicating a first-order AFM phase transition qualitatively consistent with the DFT phase diagram. To study this transition, we have determined the value of $n$ self-consistently by numerically minimizing $\varepsilon$ via Brent's method. We use the same mapping between the NNC chain and model parameters as discussed in the main text. The quantity $\Delta-Un$ that characterizes the gap (Eqn. \ref{eqn:mean-field-bands}) is plotted in Fig. \ref{fig:meanfield}. 
\newpage
\bibliography{sm_refs}